\documentclass[fleqn,usenatbib]{mnras}
\usepackage{newtxtext,newtxmath}
\usepackage[T1]{fontenc}
\usepackage{ae,aecompl}
\usepackage{graphicx}	% Including figure files
\usepackage{amsmath}	% Advanced maths commands
\usepackage{amssymb}	% Extra maths symbols
\usepackage[version=4]{mhchem}
\title[Gravitational instability in HL Tau]{ \ce{^{13}C^{17}O} suggests gravitational instability in the HL Tau disc}

\author[Booth \& Ilee]{Alice S. Booth\thanks{E-mail: pyasb@leeds.ac.uk}
and John D. Ilee\thanks{E-mail: j.d.ilee@leeds.ac.uk}
\\
School of Physics and Astronomy, University of Leeds, Leeds LS2 9JT, UK\\
}

\date{Accepted XXX. Received YYY; in original form ZZZ}

\pubyear{2020}

\begin{document}
\label{firstpage}
\pagerange{\pageref{firstpage}--\pageref{lastpage}}
\maketitle

% Abstract of the paper
% 200 word limit i think, this is only 110 so we will be fine 
\begin{abstract}
We present the first detection of the \ce{^{13}C^{17}O} $J=3$--2 transition toward the HL~Tau protoplanetary disc. We find significantly more gas mass (at least a factor of ten higher) than has been previously reported using \ce{C^{18}O} emission. This brings the observed total disc mass to 0.2~M$_{\odot}$, which we consider to be a conservative lower limit.  Our analysis of the Toomre $Q$ profile suggests that this brings the disc into the regime of gravitational instability.   The radial region of instability (50--110\,au) coincides with the location of a proposed planet-carved gap in the dust disc, and a spiral in the gas.  We therefore propose that if the origin of the gap is confirmed to be due to a forming giant planet, then it is likely to have formed via the gravitational fragmentation of the protoplanetary disc.

%%self-gravity may be the origin of this spiral arm, and (

\end{abstract}

% Select between one and six entries from the list of approved keywords.
% Don't make up new ones.

%%%%%%%%%%%%%%%%%%%%%%%%%%%%%%%%%%%%%%%%%%%%%%%%%%

%%%%%%%%%%%%%%%%% BODY OF PAPER %%%%%%%%%%%%%%%%%%

\section{Introduction}

\begin{figure*}
	\includegraphics[trim={0 0.25cm 0 0.25cm},clip, width=0.90\hsize]{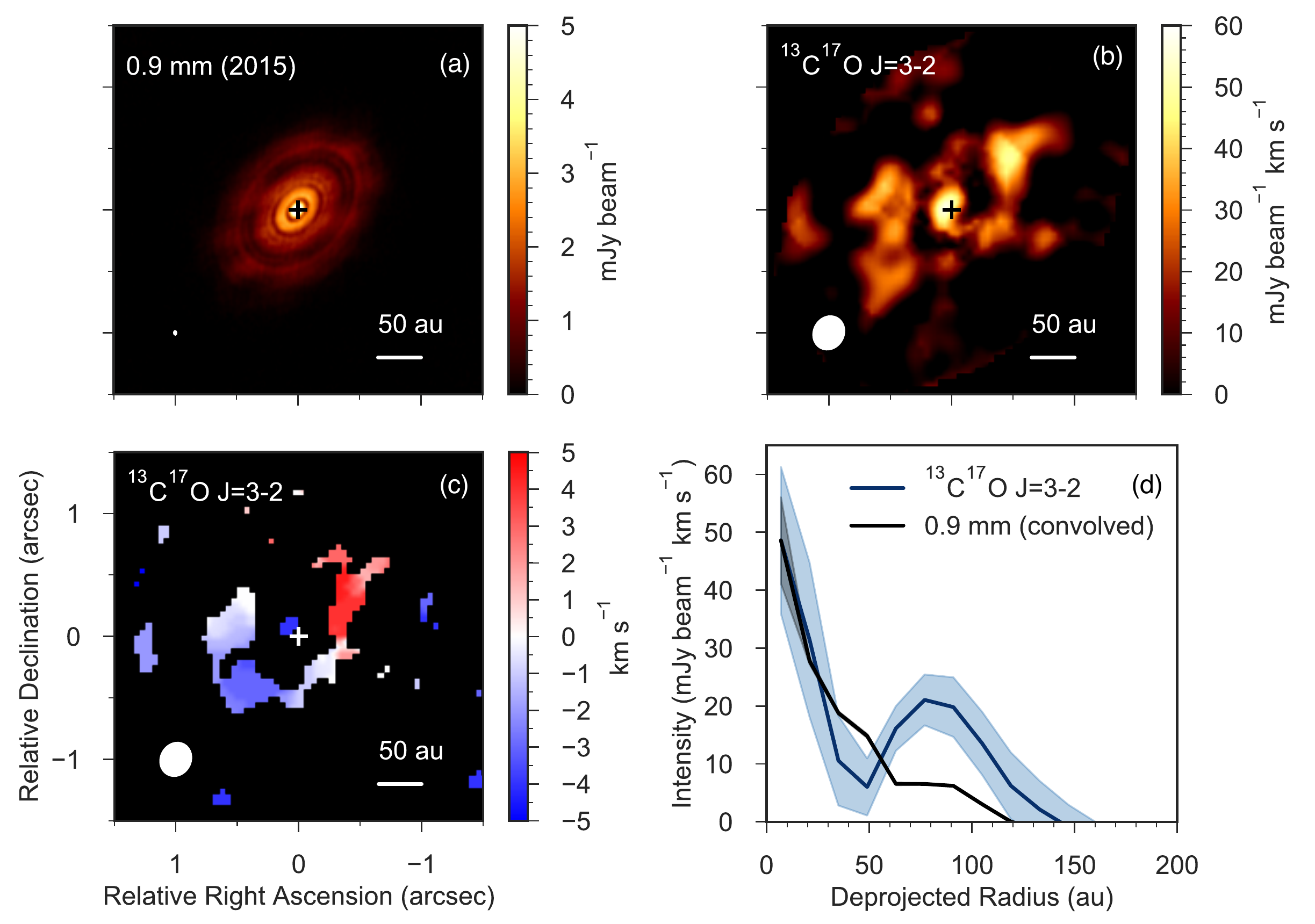}
    \caption{
\textbf{(a)} The 0.9~mm continuum image from \citet{2015ApJ...808L...3A}.
\textbf{(b)} The \ce{^{13}C^{17}O} $J=3$--2 integrated intensity map.
\textbf{(c)} The \ce{^{13}C^{17}O} $J=3$--2 intensity-weighted velocity map.
\textbf{(d)} The \ce{^{13}C^{17}O} $J=3$--2 and 0.9~mm continuum (normalised to the line emission peak) de-projected and azimuthally averaged radial profiles. The shaded regions are the errors given by the standard deviation of intensity points in each bin (0\farcs1) per beam per annulus}
    \label{maps}
\end{figure*}

The early phases of protoplanetary disc evolution set the initial conditions for planet formation. There is increasing evidence that the processes that lead to planet formation happen quickly \citep[e.g. the growth of dust grains;][]{harsono_2018}, and there is now evidence for fully-assembled protoplanetary systems in young discs \citep[$\sim$1\,Myr;][]{clarke_2018, flagg_2019}.  Characterising the properties of these discs in detail is essential if we are to understand the early evolution of young stars and systems of planets. 

\smallskip 

HL~Tau is located in the nearby (140~pc) Taurus-Auriga star forming region \citep{rebull_2004}.  Analysis of optical spectra classifies HL~Tau as spectral type K5$\pm1$ \citep{white_2004}, and spectral energy distribution (SED) modelling shows it to be a Class I--II protostar surrounded by both a circumstellar disc and envelope \citep{robitaille_2007}. The estimated visual extinction toward HL~Tau is large, with $A_{V} >  24$ \citep{close_1997}.  It also appears to be accreting at a high rate ($\dot{M} \sim 10^{-7}$~M$_{\odot}$\,yr$^{-1}$, \citealt{beck_2010}).  This observational evidence makes HL~Tau an example of an embedded, disc-hosting star with a young age \citep[$\sim1$~Myr,][]{briceno_2002}.

\smallskip

%In particular, \citet{2008MNRAS.391L..74G} targeted the HL~Tau region with the Very Large Array (VLA) at 1.3\,cm, detecting a compact feature approximately 55\,au away from the central star, suggested to be a candidate massive protoplanet forming within the disc\footnote{We note, however, that the persistence of this feature across multiple observational epochs is still under debate, see \citet{2009ApJ...693L..86C, 2019ApJ...883...71C}}. 

HL~Tau has therefore become a prime target for studies aiming to characterise young discs \citep{2008MNRAS.391L..74G, 2009ApJ...693L..86C}.  In particular, HL~Tau was the first circumstellar disc observed with the long baselines of Atacama Large Millimetre/sub-millimetre Array (ALMA).  This revealed an ordered series of concentric rings and gaps in the disc across multiple millimetre wavelengths \citep[2.9~mm, 1.3~mm, 0.9~mm;][]{2015ApJ...808L...3A}.  Many theories have been put forward to explain the origin of these structures \citep[e.g.][]{2015ApJ...806L...7Z, 2016ApJ...821...82O}, but one of the most persistent involves the growth of planets within the disc \citep{dipierro_2015_hltau}.  While efforts to detect thermal emission from young planets in the gaps have resulted in non-detections, the corresponding upper limits on their masses still lie at the higher end of the giant planet regime \citep[10--15~M$_{\rm Jup}$;][]{2015ApJ...812L..38T}.  

\smallskip

The combination of the potential for massive planets at large radii, coupled with evidence of a high disc mass \citep[$\sim$0.1--0.13\,M$_{\odot}$;][]{guilloteau_2011, 2011ApJ...741....3K} has seen gravitational instability (GI; \citealt{1997Sci...276.1836B}) being invoked to explain both the observed dust structures in HL Tau \citep{2016ApJ...818..158A, takahashi_2016}, along with the origin of any planetary companions \citep{nero_2009}. However, recent measurements of the gas mass in the HL~Tau disc using \ce{C^{18}O} have shown appears to be too low for GI to occur \citep[$M_{\rm gas} = 2$--$40\times10^{-3}$\,M$_{\odot}$;][]{2018ApJ...869...59W}. 

\smallskip

In this Letter we present a detection of the \ce{^{13}C^{17}O} $J=3$--2 transition toward the disc around HL~Tau. We have demonstrated this line to be a robust tracer of disc gas mass \citep{Booth_2019}. We use these data to derive a lower limit to the total gas disc mass and discuss how this impacts the gravitational stability of the HL Tau star-disc system.

\section{Observations}

HL Tau was observed by ALMA in Band 7 on the 24th of November 2017 for 1.46 hours with 49 antennae in configuration C43-8 under project code 2017.1.01178.S (P.I. E.~Humphreys).  Baseline lengths ranged from 92--8547~m, and the spectral window (SPW) covering the \ce{^{13}C^{17}O} $J=3$--2 transition (321.852~GHz) had a native spectral resolution of 0.908~km~s~$^{-1}$ (which we note does not resolve the $J=3$--2 hyperfine structure). 

Data (self-)calibration and imaging were performed with CASA version 5.1.1 \citep{mcmullin_2007}.  Continuum subtraction was performed in the $uv$-plane using a constant baseline (fitorder 0).  Line imaging was performed with the \textsc{clean} algorithm, using a Keplerian mask with the measured position and inclination angles of the disc \citep[e.g. ${\rm PA}=138$\degr\ and $i=46$\degr;][]{2015ApJ...808L...3A}. The native resolution of the data resulted in a natural beam size of 0\farcs14 $\times$ 0\farcs11~($-13$\degr). We used a 0\farcs2 $uv$-taper, which was found to maximise the signal-to-noise (S/N) of the line emission, resulting in a final beamsize of 0\farcs28 $\times$ 0\farcs24~($-36$\degr). The emission is detected across 9 (1.0~km~s~$^{-1}$) channels with a rms of 4~mJy~beam$^{-1}$ per channel measure from the line-free channels and a peak line emission of 30~mJy~beam$^{-1}$ per channel (S/N of 7.5).

\section{Results}
\label{sec:results}

Figure~\ref{maps} shows the ALMA Science Verification Band 7 ($\sim$345~GHz) continuum image alongside our \ce{^{13}C^{17}O} $J=3$--2 Keplerian-masked integrated intensity (zeroth moment) map and the \ce{^{13}C^{17}O} $J=3$--2 intensity-weighted velocity (first moment) map made with a 2.5~$\sigma$ clip.  Also shown are the de-projected and azimuthally averaged \ce{^{13}C^{17}O} $J=3$--2 integrated intensity and 0.9~mm continuum profiles derived from our dataset, binned to the same radial resolution.

\smallskip

The \ce{^{13}C^{17}O} emission is detected out to approximately 140\,au from the central star, with a velocity pattern that consistent with observations of other gas tracers \cite[e.g.][]{2015ApJ...808L...3A}.  There appears to be a deficit of emission (seen in both the integrated intensity map and radial emission profile) at approximately 50~au.  This feature is not coincident with any substructure  in the dust continuum or molecular gas \citep[e.g. CO, \ce{HCO^{+}};][]{yen_2016_gaps} observed at higher spatial resolution than our data. Multi-wavelength continuum observations have shown that the optical depth of the HL Tau disc at frequencies comparable to the \ce{^{13}C^{17}O} $J=3$--2 transition ($\sim$330~GHz) is high, with $\tau \sim 8$ between 40--50\,au \citep{2019ApJ...883...71C}.  This deficit of emission is therefore likely due to the high continuum opacity at Band 7 obscuring the line emission originating from the midplane, rather than a bona-fide gap in the disc.  The central peaking of the \ce{^{13}C^{17}O} profile may indicate that the line emission at radii $<10$\,au becomes optically thick at a higher height in the disc atmosphere than the dust at this frequency.  Beyond approximately 70\,au, the intensity profile follows the expected  radially-decreasing power law trend.

\begin{figure}
	\includegraphics[trim={0 0.6cm 0 0.5cm},clip,width=0.9\columnwidth]{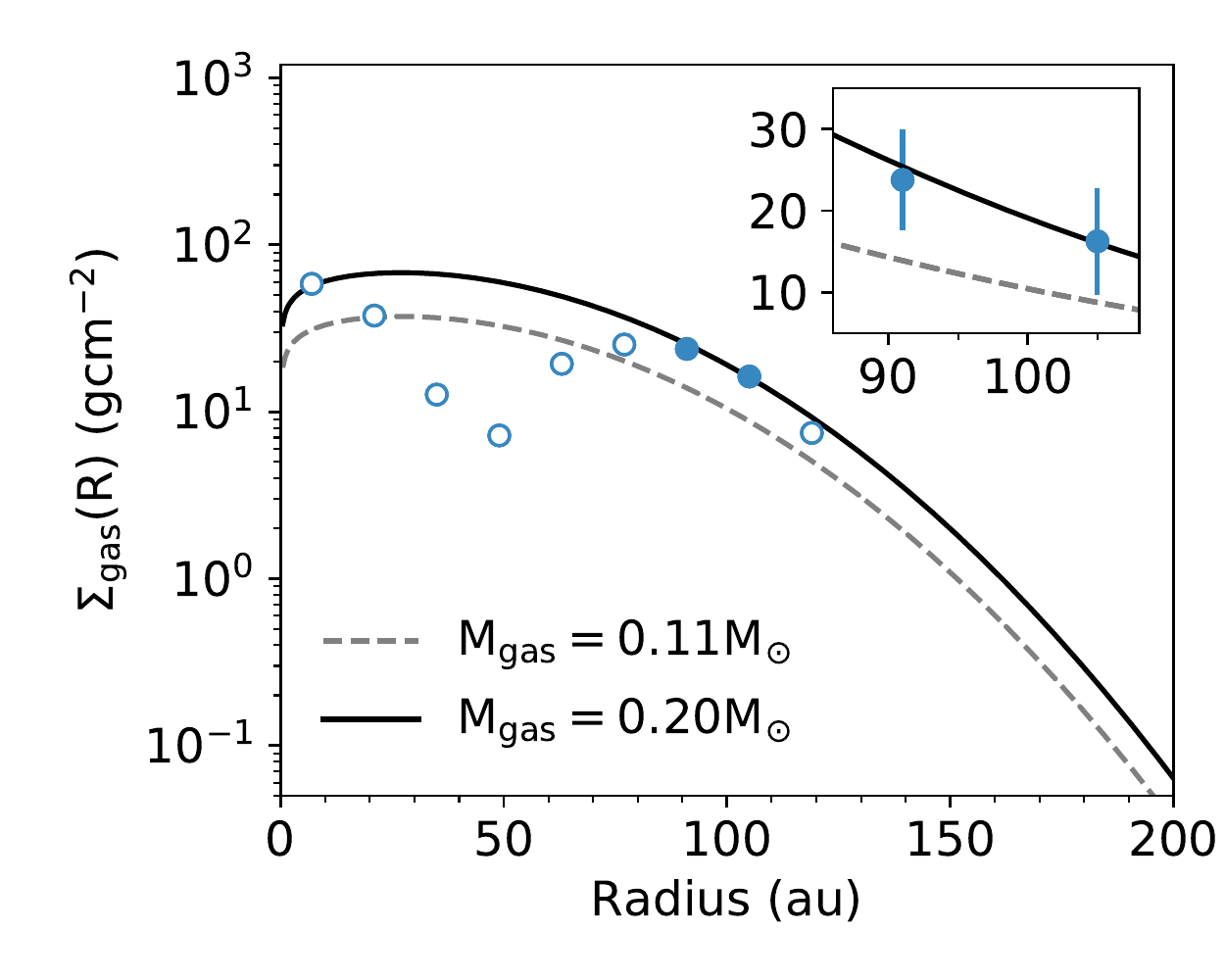}
    \caption{Comparison of total gas surface density derived from the \ce{^{13}C^{17}O} integrated intensity profile (circles) with our initial and final disc model (dashed and solid lines, respectively). Filled circles show the points used to fit the model, which the inset shows in more detail.  Errors are propagated from the observations. }
    \label{model}
\end{figure}

\section{Analysis}

\subsection{Conversion of line flux to disc gas mass}

We use the \ce{^{13}C^{17}O} observations in order to indirectly measure the total gas mass of the disc. In our previous study of the HD~163296 disc we performed 2D radiative transfer models of multiple \ce{CO} isotopologues and transitions using a well tested 2D physical structure \citep{Booth_2019}. Many molecular line observations toward HL~Tau are significantly affected by absorption and complex velocity gradients from the surrounding cloud \citep[see, e.g.,][]{2015ApJ...808L...3A, yen_2019_spiral}, meaning a similar wealth of observational data is not readily available.  We therefore employ a simpler method of analysis.  
\smallskip

The HL~Tau mm-dust surface density distribution has been shown to follow a power-law distribution:
\begin{equation}
\Sigma(r) = \Sigma_0 ~\left(\frac{r}{r_c}\right)^{- \gamma} \exp \left(-\frac{r}{r_c}\right)^{2-\gamma}.     
\end{equation}
with,
\begin{equation}
  \Sigma_0 =  (2-\gamma) \frac{M_{\rm d}}{2 \pi r_c^2} \exp{\Big( \frac{r_{\rm in}}{r_c} \Big)^{2-\gamma}} 
\end{equation}
where the total disc mass is $0.105~M_{\odot}$ (assuming a gas-to-dust mass ratio of 100), $R_c$ is 80~au, $R_{in}$ is 8.78~au and $\gamma$ is -0.20 \citep{2015ApJ...808..102K}.  This model is shown in Figure~\ref{model} as the grey dashed line. Although the observed emissivity profiles of the millimetre and centimetre sized dust are more complex, the above profile has been shown to be consistent with these \citep{2016ApJ...816...25P} and there is no evidence that these features are also present in the bulk gas distribution.  We therefore utilise this profile in our analysis. 

% Just keeping this in case we need it elsewhere... 
% We note that other surface density profiles for the HL Tau disc exist, these have greater values of \gamma (0.5-1.5; e.g. Okuzumi 2016) which will primarily lead to an increase in mass in the inner disk. Our observations are not sensitive to this radial region and thus adopting a steeper \gamma value will just lead to an increase in mass, in a reservoir we are not tracing.

\smallskip

A range of midplane temperature profiles exist for the HL~Tau disc \citep[e.g.][]{2015ApJ...808..102K, 2015ApJ...806L...7Z, 2016ApJ...821...82O, 2019ApJ...883...71C}. The model from \citet{2015ApJ...808..102K} is consistent with a $T_{\rm mid} = 55\,{\rm K} (r/10~{\rm au})^{-0.65}$ for $r>10$~au. While this is colder than other temperature structures \citep[e.g.][]{2015ApJ...806L...7Z}, warmer models have no CO snowline ($\sim 20$\,K) within the radial extent of the disc.  This is in tension with recent observations of a chemical tracer of the CO snowline,  \ce{N_2H^+}, which has been detected in the HL~Tau disc (C.~Qi, priv. comm.).  We therefore utilise the above temperature structure, which places the midplane CO snowline at approximately 50~au. 

\smallskip

We convert the observed integrated line intensity of \ce{^{13}C^{17}O} to a column density of \ce{^{13}C^{17}O} under the assumption of both local thermodynamic equilibrium (LTE) and optically thin line emission.  For each point in the radial profile we calculate 
\begin{equation}
N({\ce{^{13}C^{17}O}})=2.04 \frac{\int I_{v} d v}{\theta_{\mathrm{a}}\theta_{\mathrm{b}}} \frac{Q_{\mathrm{rot}} \exp (E_{\mathrm{u}} / T_{\mathrm{ex}})} {v^{3} \langle S_{\mathrm{ul}}\mu^{2} \rangle} \times 10^{20} \mathrm{cm}^{-2},
\label{eqn:column}
\end{equation}
where $\int I_{\nu} dv$ is the integrated line intensity in $\mathrm{Jy~beam^{-1}}$ $\mathrm{km~s^{-1}}$, $\theta_a$ and $\theta_b$ are the semi-major and semi-minor axes of the synthesized beam in arcseconds, $T_{\rm ex}$ is the excitation temperature in K, and $\nu$ is the rest frequency of the transition in GHz \citep{2003ApJ...590..314R}. The partition function ($Q_{\rm rot}$), upper energy level ($E_u$, in K), and the temperature-independent transition strength and dipole moment ($S_{\rm ul}\mu^2$, in debye$^{2}$) are taken from the CDMS database \citep{2005JMoSt.742..215M}.

\smallskip

Since the innermost regions of the \ce{^{13}C^{17}O} emission are affected by continuum absorption, we concentrate our efforts toward reproducing the level of emission between $\sim$90--110~au. As this radial location is beyond the model midplane CO snowline, any emission will be tracing CO gas from the molecular layer down to the CO snow surface \citep[see][]{miotello_2014}.  We therefore adopt $T_{\rm ex} = 25$~K for the excitation temperature in this region \citep[see][]{2012arXiv1205.6573A}, but note that a range of values for this parameter (20--80~K) result in values of column density that are well within the observed error in the integrated intensity profile. 

\begin{figure*}
% 	\includegraphics[trim={0 0.25cm 0 1.4cm},clip,width=0.47\hsize]{toomre_nov.pdf}
% 	\hspace{0.75cm}
% 	\includegraphics[trim={0 0.25cm 0 0.8cm},clip,width=0.435\hsize]{hltau_unstable.pdf}
 	\includegraphics[trim={0 0.3cm 0 0.3cm},clip,width=0.9\hsize]{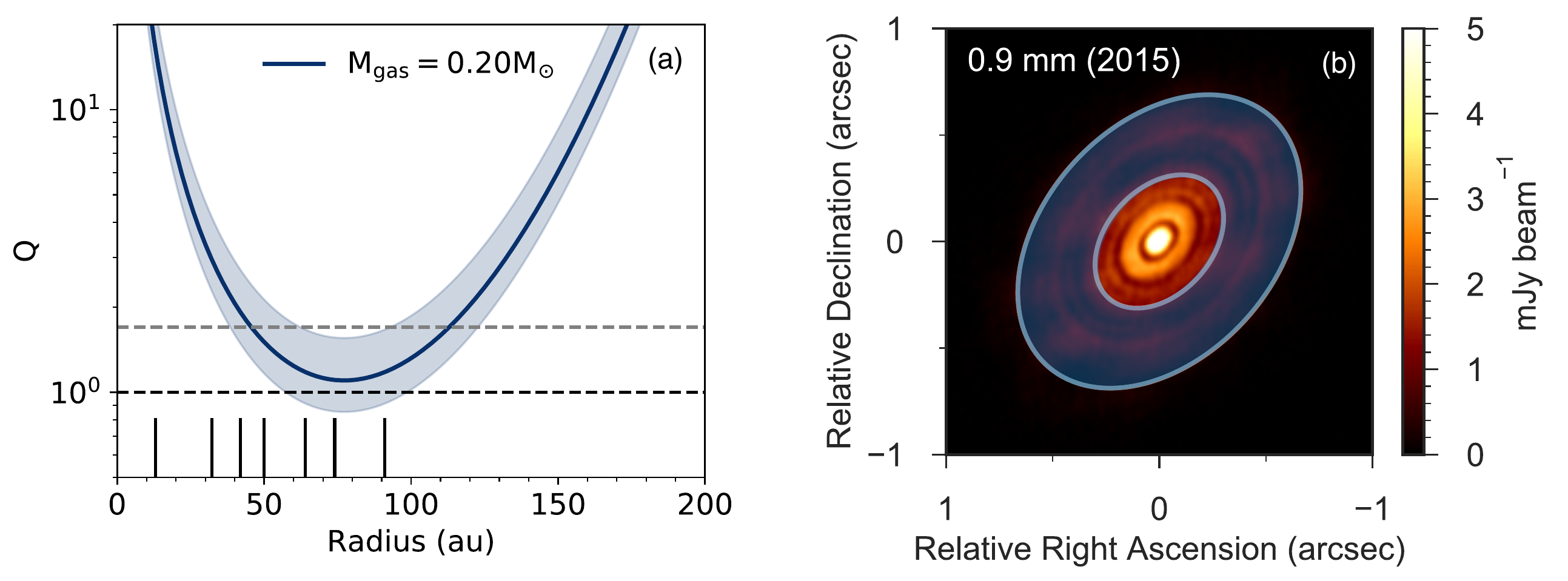}
    \caption{
    \textbf{(a)} Radial Toomre $Q$ parameter for our derived disc mass. Shaded regions denote the errors propagated from the radial intensity profile. Dashed grey and black lines mark the $Q=1.7$ and $Q=1.0$ levels, respectively. The radial locations of the `D' gaps reported in \citet{2015ApJ...808L...3A} are shown with vertical black ticks. \textbf{(b)} Same as Figure \ref{maps}(a), where the blue shaded area highlights the unstable region of the disc, with $Q<1.7$ from (a).}
    \label{Qplots}
\end{figure*} 

\smallskip

To convert from \ce{^{13}C^{17}O} to \ce{^{12}C^{16}O} we assume isotope ratios consistent with the ISM, 
$n(\ce{^{12}C^{16}O})/n(\ce{^{13}C^{16}O}) = 69$,
$n(\ce{^{12}C^{16}O})/n(\ce{^{12}C^{18}O}) = 557$, and 
$n(\ce{^{12}C^{18}O})/n(\ce{^{12}C^{17}O}) = 3.6$ \citep{Wilson_1999}.
%
%\begin{gather*}
%n(\ce{^{12}C^{16}O})/n(\ce{^{13}C^{16}O}) = 69,\\
%n(\ce{^{12}C^{16}O})/n(\ce{^{12}C^{18}O}) = 557,\\
%n(\ce{^{12}C^{18}O})/n(\ce{^{12}C^{17}O}) = 3.6.
%\end{gather*}
%
We then extrapolate to a \ce{H_2} column density assuming a moderately-depleted disc averaged $n$(\ce{^{12}C^{16}O})/$n$(\ce{H_2}) ratio of $5\times10^{-5}$.  A total gas density is then calculated assuming 80\% of the gas mass lies in \ce{H_2} \citep[e.g][]{2013ApJ...774...16R}. 

\smallskip

Based on the above, we find that an increase in the total disc gas mass to $M_{\rm gas} = 0.20^{+0.06}_{-0.06}$~M$_{\odot}$ is required to match the observations between 90--110~au (where errors are propagated from the observed radial intensity profile). The resulting curve is shown in Figure~\ref{model} alongside our gas surface density values derived from the radial intensity profile of \ce{^{13}C^{17}O}.  

\subsection{Assumptions influencing the derived gas mass}

The conversion of line flux to gas mass depends on several
assumptions, which we discuss here.  Our derived gas mass is sensitive to the chosen conversion factor between CO and \ce{H_2} \citep[see][for a review]{2018arXiv180709631B}.  While this factor is often taken to be $\sim$10$^{-4}$, studies of individual protoplanetary discs have revealed carbon depletion from factors of five to orders of magnitude \citep[e.g.][]{McClure_2016}, and surveys have demonstrated this phenomenon is widespread \citep[e.g.][]{miotello_2017}.  Our chosen value of $n$(\ce{^{12}C^{16}O}$)/n$(\ce{H_2}$) = 5\times10^{-5}$ therefore accounts for a modest depletion factor of two, in agreement with disc chemical models at ages of $\sim$1\,Myr \citep{Schwarz_2018, 2018A&A...618A.182B}\footnote{Though we note depletion factors depend on variables such as temperature and ionisation rate.}. The derived disc gas mass scales linearly with the inverse of the depletion factor, so under the assumption of no carbon depletion (e.g. 10$^{-4}$), our derived disc mass would be half our quoted value.

\smallskip

We have also assumed that our observations trace the full column of CO on both the `near' and `far' sides of the disc.  This will not be the case if $i$) there is a significant fraction of CO frozen out in the disc midplane, or if the \ce{^{13}C^{17}O} $J=3$--2 emission is optically thick.  In addition (as discussed in Section \ref{sec:results}), \citet{2019ApJ...883...71C} have determined that the continuum optical depth in the disc is high, with $\tau \sim 8$ at 330\,GHz.  Therefore, our observations are not sensitive to CO emitting on the far side of the disc.  In this scenario, the true gas mass in the disc would be (at least) a factor of two higher.   

\smallskip

Observations of \ce{CO} rovibrational lines toward the HL~Tau envelope have revealed a significantly higher n(\ce{^{12}C^{16}O})/n(\ce{^{12}C^{18}O}) ratio of $760~\pm~80$, which is consistent with isotope-selective photo-dissociation and self-shielding \citep{2009A&A...503..323V, 2015ApJ...813..120S}. But, as these higher energy infrared transitions do not trace the same reservoir of CO probed in our sub-mm observations, we adopted the ISM values. If the higher ratio applies, then a higher column of CO would be required to match our observations, increasing our derived gas mass (though by factors lower than the above examples).

\smallskip 

The precise magnitude of these effects cannot be determined without dedicated modelling of the HL Tau disc.  Nevertheless, we believe our estimates for these factors imply that our derived gas mass is a conservative lower limit. Future high spatial and spectral resolution observations of multiple CO isotopologues toward HL~Tau will enable us to place more stringent constraints on the gas mass in the disc.

\section{Discussion}

\subsection{Comparison to other mass measurements}

\citet{tapia_2019} use combined dust and gas evolutionary disc models to reproduce ALMA and VLA continuum profiles of the HL Tau disc from to 0.87--7.8\,mm.  Their best fitting model possesses a dust mass of $4.8\times10^{-3}$\,M$_{\odot}$ with a gas-to-dust mass ratio of 50, resulting in a disc gas mass of 0.23\,M$_{\odot}$.  Our result from \ce{^{13}C^{17}O} therefore brings mass measurements from dust continuum and molecular line emission into agreement for the first time in this disc.

\smallskip

\citet{2018ApJ...869...59W} use \ce{C^{18}O} $J=2$--1 SMA observations of HL~Tau to derive a total disc gas mass of 2--$40\times10^{-4}$~M$_{\odot}$, which is between a factor of 50--1000 smaller than the gas mass implied by our results.  However, the above work adopts a CO/\ce{H_2} ratio of $2.7\times10^{-4}$, which assumes no C depletion, and that all potential volatile carbon in the disk is the form of CO (given an ISM C/H ratio of $1.4\times10^{-4}$; \citealt{1996ApJ...467..334C}). If we adopt the same ratio as \citet{2018ApJ...869...59W}, our resulting gas mass would be $\sim$0.04~M$_{\odot}$, which is still a factor of 10--200 higher than the gas mass derived from \ce{C^{18}O}.

\smallskip

Using Equation \ref{eqn:column} we can compare the values for N(\ce{^{13}C^{17}O}) and N(\ce{^{12}C^{18}O}).   A column density ratio of $\sim$250 would imply that both lines are optically thin.  However, using our data and the data from the radial profile presented in \citep{2018ApJ...869...59W} for a radius of 100~au, we find that the ratio is only $\sim 4$.  This demonstrates that the \ce{C^{18}O} $J=2$--1 emission is optically thick in the HL~Tau disc, and is therefore not tracing the bulk gas mass in the molecular layer \citep[as also found for HD~163296;][]{Booth_2019}.

% dust mass from Tapia - 4.8×10−3M⊙ within 100 au
% dust mass from CG 2016 - (1-3)×10−3M⊙
% using Tapia of g/d ratio is 0.20/0.005 approx 40
% we do  match Tapia within our mass error range 

\subsection{The (in)stability of the disc}

The gravitational stability of the higher mass disc can be assessed using the Toomre~$Q$ parameter \citep{1964ApJ...139.1217T}, 
\begin{equation}
     Q = \frac{c_{s} \kappa}{\pi G \Sigma},
\end{equation}
where $c_s$ is the sound speed of the gas (calculated from T$_{\rm mid}$), $\Sigma$ is the surface density of the gas and dust, and $\kappa$ is the epicyclic frequency (equal to the angular velocity $\Omega$ in a Keplerian disc where we assume a central star mass of 1.7~M$_{\odot}$, \citealt{2016ApJ...816...25P}).  The resulting radial profile of $Q$ is shown in Figure~\ref{Qplots}a.   

\smallskip

Regions of the disc with $Q \lesssim 1.7$ will be susceptible to gravitational instability \citep{2007prpl.conf..607D, helled_2014}, leading to non-axisymmetric structure in the disc and potentially fragmentation.  Our observations therefore support a picture in which the disc around HL~Tau is gravitationally unstable ($Q<1.7$) from approximately 50--110\,au (with minimum  $Q = 1.1_{-0.2}^{+0.5}$ at $r  = 77$\,au).  Taking into account observational uncertainties, it is also possible that $Q<1$ between 60--100\,au.

\smallskip

If the disc around HL~Tau is threaded by a magnetic field, then this may offer an additional mechanism to support the disc against self-gravity.  In these cases, $Q$ is modified by a factor $\sqrt{1 + 1/\beta_{\rm p}}$ (where $\beta_{\rm p}$ is the plasma parameter) and the threshold for instability lies within the range $Q \lesssim1.2$--1.4 \citep{kim_2001}.  In general, values of $\beta_{\rm p}$ range from $\sim$10 under ideal MHD conditions \citep{2017MNRAS.466.3406F} up to 10$^{4}$ for non-ideal MHD \citep{2017ApJ...845...31H}.  Such values do not alter our minimum Toomre~$Q$ value by more than $\sim$5 per cent, suggesting that even if the HL Tau disc possesses a strong magnetic field ($\beta_{\rm p} \sim 10$), it would not be sufficient to move the disc into a stable regime. 

\subsection{The region of instability in context}

Figure \ref{Qplots}b shows the region of instability in the HL~Tau disc with respect to the Band 7 (0.9\,mm) continuum observations of \citet{2015ApJ...808L...3A}.  The unstable region spans several of the bright rings and dark gaps identified in the dust disc (B3--B7 and D4--D7). Modelling of unstable discs with a decoupled dust component shows grains concentrating in spiral arms \citep[e.g.][]{rice_2004, dipierro_2015_spirals}. Therefore it is not immediately clear how to reconcile such an apparently unstable disc with the ordered concentric rings observed in the dust.

\smallskip

Several studies have shown that the `double gap' feature from $\sim$65--74\,au (D5--B5--D6) can be reproduced by the presence of a planet with mass between $\sim$0.2--0.6\,M$_{\rm Jup}$  \citep{dipierro_2015_hltau, 2016ApJ...818...76J, 2018ApJ...866..110D}, which is supported by mm--cm observations across this radial region \citep{2019ApJ...883...71C}.
The higher gas surface density revealed by our observations implies that mm-sized grains would be better coupled to gas than in the above works.  Therefore, larger planet masses would be required to match the observed dust structures.  When considering the lower bound of our observational errors, we find $Q \leq 1$ from $\sim$60--100\,au.  Here the local cooling time is expected to be short \citep[e.g.][]{rafikov_2005, clarke_2009}, meaning that this region of the disc would undergo gravitational fragmentation.  This scenario is in agreement with dedicated modelling of star-disc systems with similar properties as HL Tau, which form fragments with masses between 1-5~M$_{\rm Jup}$ \citep{2011ApJ...731...74B}.  Therefore, if such a planet is confirmed to be the origin of this observed gap structure, then we suggest it has likely formed via gravitational fragmentation.

\smallskip

Recently, \cite{yen_2019_spiral} detected a spiral feature in observations of \ce{HCO^+} $J=3$--2 toward HL~Tau.  While they attribute the spiral to in-falling material from the surrounding envelope, this feature crosses our region of instability in the disc. Chemical models of gravitationally unstable discs have predicted \ce{HCO^{+}} to be a tracer of spiral structure \citep{ilee_2011, douglas_2013}.  In light of this, we suggest that the spiral structure on small ($\lesssim$100\,au) scales could be due to gravitational instability within the disc.

\section{Conclusions}

We have presented the first detection of \ce{^{13}C^{17}O} in the HL Tau disc, revealing a significant reservoir of previously hidden gas mass. The total disc mass is now sufficient for the disc to be considered gravitationally unstable between 50--110\,au.  This region crosses a spiral feature observed in the gas, and a proposed planet-carved gap in the dust continuum.  We suggest that if a massive planet is confirmed to be the origin of this gap, then it likely formed via gravitational fragmentation of the protoplanetary disc.

\smallskip

This work represents only the second detection of \ce{^{13}C^{17}O} in a protoplanetary disc.  It further demonstrates the utility of rare CO isotopologues in probing the physical conditions in discs, with important implications for their evolution and the formation of planets within them.

\section*{Acknowledgements}

We would like to thank Catherine Walsh, Duncan Forgan, Ken Rice, Cathie Clarke \& Charlie Qi for extremely helpful discussions during the preparation of this manuscript.  We also thank the anonymous referee for a constructive report. A.B. acknowledges the studentship funded by the Science and Technology Facilities Council of the United Kingdom (STFC) and   J.D.I. acknowledges support from the STFC under ST/ R000549/1.  This paper makes use of the following ALMA data: ADS/JAO.ALMA\#2011.0.00015.SV, ADS/JAO.ALMA\#2017.1.01178.S. ALMA is a partnership of ESO (representing its member states), NSF (USA) and NINS (Japan), together with NRC (Canada), MOST and ASIAA (Taiwan), and KASI (Republic of Korea), in cooperation with the Republic of Chile. The Joint ALMA Observatory is operated by ESO, AUI/NRAO and NAOJ.

%\bibliographystyle{mnras}
%\bibliography{hltau_2020} 

%\appendix

% %\section{Channel Maps}

% %Figure \ref{channel_maps} shows the channel maps resulting from our line imaging.  The emission is detected across 9 (1.0~km~s~$^{-1}$) channels with a $\sigma$ of 0.004~Jy~beam$^{-1}$ per channel and a peak emission of 0.030~Jy~beam$^{-1}$ per channel (S/N of 7.5).

% %\begin{figure*}
% %    \centering 
% %	\includegraphics[trim={0 0.4cm 0 1.5cm},clip, width=0.95\hsize]{hltau_13c17o_channels_kep_mask.pdf}
% %    \caption{The \ce{^{13}C^{17}O} J=3-2 channel maps at 1~km~s$^{-1}$. Contours mark the 3, 5 and 7 $\sigma$ levels where $\sigma$ is 6~mJy~beam$^{-1}$ channel$^{-1}$. The dashed black lines show the Keplerian mask used in the cleaning procedure.}
% %    \label{channel_maps}
% %\end{figure*}

% \begin{figure*}
% 	\includegraphics[trim={0 0.25cm 0 0.25cm},clip, width=0.90\hsize]{hltau_moments.pdf}
%     \caption{
% \textbf{(a)} The 0.9~mm continuum image from \citet{2015ApJ...808L...3A}.
% \textbf{(b)} The \ce{^{13}C^{17}O} $J=3$--2 integrated intensity map.
% \textbf{(c)} The \ce{^{13}C^{17}O} $J=3$--2 intensity-weighted velocity map.
% \textbf{(d)} The \ce{^{13}C^{17}O} $J=3$--2 and 0.9~mm continuum (normalised to the line emission peak) de-projected and azimuthally averaged radial profiles. The shaded regions are the errors given by the standard deviation of intensity points in each bin (0\farcs1) per beam per annulus}
%     \label{maps}
% \end{figure*}

% Don't change these lines
\bsp	% typesetting comment
\label{lastpage}
\end{document}